\documentclass[conference]{IEEEtran}

\usepackage[utf8]{inputenc} 	% UTF-8

\usepackage{amsmath}
\usepackage{amssymb}
\usepackage{amsfonts}
\usepackage{amsthm}
\usepackage{bbm}
\usepackage{multicol}
\usepackage{cite}

\newtheorem{defi}{Definition}

\usepackage[pdftex]{graphicx}
\newcommand{\e}[2]{{\mathbb E}_{\textbf{#1}} {\left[ #2 \right]}}

\newcommand{\eqdef}{\mathrel{\mathop:}=}
\ifCLASSOPTIONcompsoc
\usepackage[caption=false,font=normalsize,labelfont=sf,textfont=sf]{subfig}
\else
\usepackage[caption=false,font=footnotesize]{subfig}
\fi

\usepackage[final=true]{hyperref}
\hypersetup{
	pdfauthor = {Ankit Kaushik et al.},
	pdftitle = {WCNC 2014},
	pdfsubject = {WCNC 2014},
	pdfcreator = {PDFLaTeX with hyperref package},
	pdfproducer = {PDFLaTeX}
}

\begin{document}
\title{On the Estimation of Channel State Transitions for Cognitive Radio Systems}

\author{Ankit Kaushik, Friedrich K. Jondral \\ Communications Engineering Lab \\ Karlsruhe Institute of Technology (KIT) \\ \{\href{mailto:Ankit.Kaushik@kit.edu}{Ankit.Kaushik}, \href{mailto:Friedrich.Jondral@kit.edu}{Friedrich.Jondral}\}@kit.edu
}

% make the title area
\maketitle
\thispagestyle{empty}
\pagestyle{empty}

\begin{abstract} 
Coexistence by means of shared access is a cognitive radio application. The secondary user models the slotted primary users channel access as a Markov process. The model parameters, i.e, the state transition probabilities $(\alpha, \beta)$ help secondary user to determine the channel occupancy, thereby enables secondary user to rank the primary user channels. These parameters are unknown and need to be estimated by secondary users for each channel. To do so, the secondary users have to sense all the primary user channels in every time slot, which is unrealistic for a large and sparsely allocated primary user spectrum. %A paradigm shift will occur only when the theoretical models could also operate in practical scenarios. One such aspect is considered in the paper, 
With no other choice left, the secondary user has to sense a channel at random time intervals and estimate the parametric information for all the channels using the observed slots. %In this way secondary user could schedule the channels in every time slot by selecting by lesser number of channels to sense and still obtains the parametric information for all the channels.       

\end{abstract}
%%%%%%%%%%%%%%%%%%%%%%%%%%%%%%%%%%%%%%%%%%%%%%%%%%%%%%%%%%%%%%%%%%%%%%%%%%%%%%%%%%%%%%%%%
\section{Introduction}
%%%%%%%%%%%%%%%%%%%%%%%%%%%%%%%%%%%%%%%%%%%%%%%%%%%%%%%%%%%%%%%%%%%%%%%%%%%%%%%%%%%%%%%%%
Cognitive Radio (CR) as introduced by Mitola in \cite{Mitola} is a paradigm shift, which integrates intelligence into the radio system. Bringing cognition into the existing radio system allows the allocated spectrum to be used more efficiently \cite{Weiss}. A typical scenario is where Secondary Users (SUs) and Primary User (PUs) coexist through sharing, i.e, SUs harvest spectrum holes inside PU spectrum. Without any cooperation, each SU follows the slotted medium access made by the PU over its channels through sensing.  \\
SU considers the discrete time discrete state Markov process to model the PU channel access \cite{Gilbert}. The state transition probabilities as model parameters provide information about the channel occupancy, which enables SU to rank the PU channels and detect the spectrum holes in an efficient way. In this way, each SU could exercise Reinforcement Learning (RL) for medium access. RL constitutes of two phases: exploration and exploitation \cite{Sutton98}. In the exploration phase, SU gathers occupancy information for the PU channels and utilizes it for its transmission in the exploitation phase. \\
Markov process is relatively simple model to describe slotted medium access, although, due to its analytical tractability, it is mostly considered to perform analysis \cite{Kim2008, Zhao, Long}. \cite{Kim2008} ensures the fastest discovery of the idle channels at the SU. \cite{Zhao} describes the process of selecting the PU channels for sensing followed by their access as the multi-armed bandit problem. These approaches assume the knowledge of the model parameters at the SU, which constitutes only the exploitation phase. The exploration phase that comprises of determining the model parameters using Maximum Likelihood Estimation (MLE) for a given channel is done in \cite{Long}. According to \cite{Long}, the transition probabilities can be estimated when a channel is sensed in consecutive slots. Following this approach, the estimation of parameters for each channels requires SU to sense all the channels at each slot. That means for a given channel a slot sequence at SU is obtained by observing all slots. The slot sequence is sufficient for the estimation of the model parameters, is defined as \textit{complete dataset}. \\
However in a practical scenario, the sparse location of the PU channels and the slot duration constrain SU to sense only a limited number of channels at each slot. From the SU perspective, this is equivalent to a situation where a slot sequence for a given channel is obtained by observing non-consecutive slots, is defined as \textit{incomplete dataset}. This makes the estimation of the model parameters using MLE analytically not possible. Dempster \textit{et al.} \cite{Dempster77} proposed Expectation-Maximization (E-M) algorithm, an iterative approach to find the MLE for incomplete dataset. \\
As per authors' knowledge, E-M approach in the literature has been limited to Hidden Markov model and Gaussian Mixture model \cite{BILMES98}, hence, never applied to discrete time discrete state Markov process with incomplete dataset. E-M algorithm leads to an optimal solution, but it is applicable only to the probabilistic models with exponential families. Also, the likelihood function of the incomplete dataset has multiple modes. Thus, E-M algorithm like any other hill climbing algorithm, may not converge to a global maximum. \\
The paper formulates the problem of parameter estimation for a given channel following the Markov process when only non-consecutive slots are observed. Secondly, we derive an analytical expression of likelihood function for the Markov process as an exponential family, and apply the E-M algorithm for estimating the model parameters for the incomplete dataset. Simulations are performed to validate the E-M algorithm. Finally, we apply least squared error algorithm to find the global maximum. However to keep the discussion limited and yet focused, the paper considers perfect sensing, i.e probability of false alarm $P_{fa} = 0$ and probability of detection $P_{d} = 1$. \\
The rest of the paper is organized as follows: Section \ref{sec:chan mod} explains the system model. Section \ref{sec:MLE} presents the analytical expression for the MLE for the complete and incomplete dataset for the underlying Markov process. For the incomplete data set, it implements an iterative approach to determine MLE. The theoretical analysis is supported by Monte Carlo simulations in Section \ref{sec:sim res} followed by conclusion in Section \ref{sec:conc}. The definition of the exponential family in the natural form for the complete and incomplete dataset, and the expression of the likelihood function for Markov process as an exponential family are presented in Appendix.   
%%%%%%%%%%%%%%%%%%%%%%%%%%%%%%%%%%%%%%%%%%%%%%%%%%%%%%%%%%%%%%%%%%%%%%%%%%%%%%%%%%%%%%%%%
\section{System Model} \label{sec:chan mod}
%%%%%%%%%%%%%%%%%%%%%%%%%%%%%%%%%%%%%%%%%%%%%%%%%%%%%%%%%%%%%%%%%%%%%%%%%%%%%%%%%%%%%%%%%
\begin{figure}[!t]
	\centering
	\includegraphics{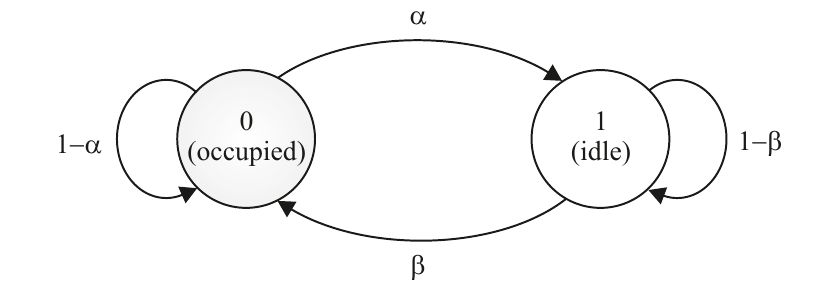}
	\caption{2-state discrete Markov channel model}
	\label{fig:ChMod}
\end{figure}
We consider $N$ PU channels. Two consecutive slots of a given channel $n$ follow a discrete state Markov chain as shown in \figurename \ref{fig:ChMod} with transition probabilities $(\alpha, \beta)\in [0,1]$ \cite{Gilbert}. Each channel is independent and has different set of parameters $(\alpha_{n}, \beta_{n})$, however to ease the notations, subscript $n$ is dropped in rest of the paper. Medium access is synchronized and slotted, i.e, the channel stays in a given state (Occupied = 0, 1 =  idle) for the complete slot duration. The probability that a given channel is in occupied state at slot, at time index $t$, is defined as the channel utilization probability $u = P(x_t = 0)$. Let $T$ be the number of observed slots for a channel, then 
\begin{equation}
	\begin{split}
	T_0 \eqdef \sum_{t=2}^{T} \mathbbm{1}_{ \{ x_{t} | x_{t - 1} = 0 \} } \\
	T_1 \eqdef \sum_{t=2}^{T} \mathbbm{1}_{ \{ x_{t} | x_{t - 1} = 1 \} }
	\end{split}
\label{eq:Suff1}
\end{equation} 
are the number of slot transitions with previous state $(x_{t-1} = 0)$ and $(x_{t-1} = 1)$, then $T_0 = T \cdot (1 - u)$ and $T_1 = T \cdot u$. $\mathbbm{1}_{\{\cdot | \cdot\}}$ represents the indicator function for a conditional event $\{\cdot|\cdot\}$. Markov process can be represented as an alternating renewal process 
\begin{align}
\label{eq:chauti1}
u := \lim_{T \rightarrow \infty} \frac{T_0}{T} = \frac{\e{}{L_0}}{\e{}{L_0} + \e{}{L_1}},
\end{align} 
where $L_0$, $L_1$ are geometrically distributed random variables denoting the number of successive slots with occupied and idle state, where $\e{}{L_0} = \frac{1}{\alpha}, \e{}{L_1} = \frac{1}{\beta}$ and $T = T_0 + T_1$. With the knowledge of transition probabilities $(\alpha, \beta)$, $u$ is evaluated as
\begin{align}
\label{eq:chauti2} 
u = \frac{\beta}{\alpha + \beta}.
\end{align} 
After obtaining $u$ for $N$ channels, i.e., $\textbf{u} = [u_{1}, u_{2},...,u_{N}]$, SU ranks the channels and optimizes its capacity.
This knowledge is not available at SU and is estimated through sensing.
%%%%%%%%%%%%%%%%%%%%%%%%%%%%%%%%%%%%%%%%%%%%%%%%%%%%%%%%%%%%%%%%%%%%%%%%%%%%%%%%%%%%%%%%%
\section{Estimation Algorithm} \label{sec:MLE}
%%%%%%%%%%%%%%%%%%%%%%%%%%%%%%%%%%%%%%%%%%%%%%%%%%%%%%%%%%%%%%%%%%%%%%%%%%%%%%%%%%%%%%%%%
%%%%%%%%%%%%%%%%%%%%%%%%%%%%%%%%%%%%%%%%%%%%%%%%%%%%%%%%%%%%%%%%%%%%%%%%%%%%%%%%%%%%%%%%%
\subsection{Maximum Likelihood Estimation for a complete dataset} \label{subsec:CData MLE}
The slot sequence of a single PU channel is represented as a random vector $\textbf{X}$, where each slot $X_t$ is a binary random variable. The estimation of the model parameters ${\theta} = (\alpha, \beta)$ from an observed data set $\textbf{x} = [x_1,x_2,...,x_t,...,x_T]$, where  $T \in \mathbb{N}$ is done using Maximum Likelihood Estimation \cite{Kim2008}. \\
Definition \ref{df:def1} in Appendix describes the nomenclature of the exponential family. Following it, Definition \ref{df:def2} represents the Markov process characterized by likelihood function $p( \textbf{x} | {\theta})$ as an exponential family.
Consider (\ref{eq:expfamily_Markov_ergodic}) and (\ref{eq:expfamily_Markov}) from Definition \ref{df:def2}, the MLE of parameters describing the Markov process are determined as
\begin{align}
	\label{eq:Suf_stat}
	\begin{bmatrix}
		\widehat{\alpha} & \widehat{\beta} 
	\end{bmatrix}	
		&=		
	\begin{bmatrix} 
		\frac{t_{01}}{T_0} & \frac{t_{10}}{T_1}
	\end{bmatrix},
\end{align}
where $t_{01}, t_{10}$, $T_0$ and $T_1$ are the sufficient statistics of a two state Markov process. $t_{01}$ and $t_{10}$ are number of consecutive slots with sample values $(x_{t-1} = 0, x_{t} = 1)$ and $(x_{t-1} = 1 ,x_{t} = 0)$ is defined as
\begin{equation}
	\begin{split}
		t_{01} \eqdef \sum\limits_{t=2}^{T} \mathbbm{1}_{ \{ x_t = 1 | x_{t-1} = 0  \}} \\
		t_{10} \eqdef \sum\limits_{t=2}^{T} \mathbbm{1}_{ \{ x_t = 0 | x_{t-1} = 1  \}}.
	\end{split}
\label{eq:Suff2}
\end{equation}
$T_0$ and $T_1$ already are defined in (\ref{eq:Suff1}). The sufficient statistics are evaluated after observing the sequence \textbf{x}.

%%%%%%%%%%%%%%%%%%%%%%%%%%%%%%%%%%%%%%%%%%%%%%%%%%%%%%%%%%%%%%%%%%%%%%%%%%%%%%%%%%%%%%%%%
\subsection{Maximum Likelihood Estimation for an incomplete dataset}\label{subsec:ICData MLE}
\begin{figure}[t!]
	\centering
	\includegraphics[trim=0.0cm 0.0cm 0.0cm 0.0cm,clip=true,width=\columnwidth]{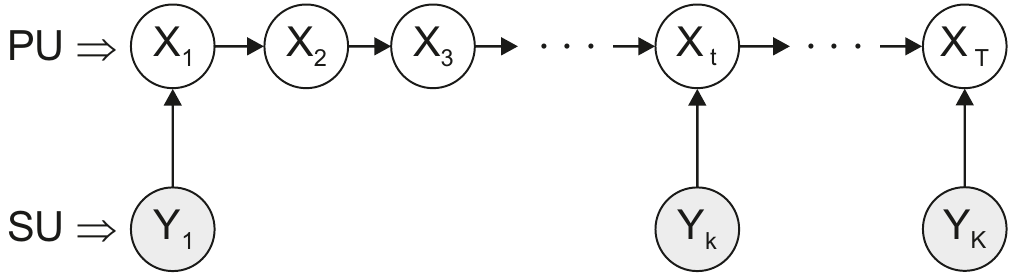}
	\caption{A state sequence \textbf{X} of a primary user channel following a Markov process, observed by a secondary user denoted as \textbf{Y}.}
	\label{fig:Incomdataseq}
\end{figure}
Until now, the parameters for each channel are estimated assuming that all channels are sensed at each of the $T$ slots. But in a real scenario, $N$ is large and the channels are sparsely distributed in the spectrum. The sampling rate of A/D converters and time slot duration bound the number of channels that can be sensed in a given time slot by SU. \\
At the start, SU has no knowledge of the parameters, therefore it randomly selects $M (< N)$ channels for sensing in each slot. Without loss of generality, this is equivalent to the case when a single channel is sensed by the SU at $K$ non-consecutive slots. Now, the task of SU is to estimate the parameters for the channel from $K$ observed slots. Therefore, we consider only a given PU channel that follows a slot sequence $x_T$, and $y_K$ are the slots observed by SU as described in \figurename \ref{fig:Incomdataseq}, where $K \le T$, with equality when the SU observes all the slots. \\   
The SU observing a single PU channel illustrated in \figurename \ref{fig:Incomdataseq} is represented as
\begin{equation}
	\label{eq:Incompl_Representation}
	\textbf{Y} = h(\textbf{X}), 
\end{equation}
where $h:\mathcal{X}^{T} \rightarrow \mathcal{Y}^K$ is many to one mapping. ${(\cdot)}^{(\cdot)}$ represents the Cartesian product. $\textbf{X} \in \mathcal{X}^T$ is a sequence of slots that symbolizes the complete dataset and $\textbf{Y} \in \mathcal{Y}^K$ are the slots that are observed by the SU represents the incomplete dataset with probability measures $p(\textbf{x} | {\theta})$ and $p(\textbf{y} | {\theta})$. The slot indices for the PU and SU are denoted by $t$ and $k$, see \figurename \ref{fig:Incomdataseq}. The information gained after observing $\textbf{y}$ confines the sample space $\mathcal{X}^T$ to a sub-space $\mathcal{X}^T(\textbf{y}) \subset \mathcal{X}^T$, with probability measure $p(\textbf{x} | \textbf{y}, {\theta})$. The relationship between probability measure for the incomplete and complete data is given by 
\begin{equation} 
\label{eq:Like_incom_probrel}
p(\textbf{y}|{\theta}) = \sum_{\mathcal{X}^T(\textbf{y})} p(\textbf{x}|{\theta}).
\end{equation}
We consider $T = LK + 1$, where $L$ is the skipped slots length, i.e., the SU observes a given PU channel after skipping $L$ slots. Now, $L$ = constant depicts the periodic behavior for slot observation, where as random value of $L$ refers to a general case. \\
Without loss of generality, we consider first the periodic case where $L$ is fixed and then generalize to the random case. \figurename \ref{fig:Incomdataseq} shows that, when ${(k)}^{th} = {(t)}^{th}$ then ${(k + 1)}^{th} = {(t + L + 1)}^{th}$.
For the case where $T = 5, K  = 2$, $L = 3$ and $y_1 = 0, y_5 = 1$, $\mathcal{X}^T(\textbf{y}) = [0,x_2,x_3,x_4,1]$ corresponds to $2^L$ combinations of path sequences $\{ [0,0,0,0,1], [0,0,0,1,1],..., [0,1,1,1,1] \}$. Applying Markov property to define $p(\textbf{y}|{\theta})$ in terms of conventional parameters $(\alpha, \beta)$ over the sample space $\mathcal{X}^T(\textbf{y})$
\begin{equation} 
\label{eq:Like_incom_probrel_ex}
\begin{split}
p(\textbf{y}|{\theta}) =& (1 - \alpha)^3 \cdot \alpha + (1 - \alpha)^2 \cdot \alpha \cdot (1 - \beta) +  \\ \quad & (1 - \alpha) \cdot \alpha^2 \cdot \beta + (1 - \alpha) \cdot \alpha \cdot (1 - \beta)^2 +  \\ \quad &  (1 - \alpha) \cdot \alpha^2 \cdot \beta +  \alpha^2 \cdot (1 - \beta) \cdot \beta + \\ \quad &  \alpha^2 \cdot (1 - \beta) \cdot \beta +  \alpha \cdot (1 - \beta)^2. 
\end{split}
\end{equation}
Clearly, the expression $p(\textbf{y}|{\theta})$ given in (\ref{eq:Like_incom_probrel_ex}) forms a non-linear expression in terms of $(\alpha,\beta)$, which cannot be represented as an exponential family, as described in Definition \ref{df:def1}. However, under the Bayesian framework $p(\textbf{y}|{\theta})$ can be represented as
%\begin{equation} 
\begin{align}
	\label{eq:Like_incom1}
	 p(\textbf{y}|{\theta}) &= p(\textbf{x}, \textbf{y} | {\theta}) / p(\textbf{x} | \textbf{y},{\theta}), \nonumber \\
	 \intertext{inserting $p(\textbf{x}, \textbf{y} | {\theta}) = p(\textbf{x} | {\theta})$ and taking log delivers} 
	% p(\textbf{y}|{\theta}) &= p(\textbf{x} | {\theta}) /  p(\textbf{x} | \textbf{y},{\theta}), \nonumber 
	%  \intertext{Taking log} 
	 \log  p(\textbf{y}|{\theta})  &= \log p(\textbf{x} | {\theta}) - \log p(\textbf{x} | \textbf{y},{\theta}).
\end{align}
Maximizing $\log p(\textbf{y}|{\theta}) $ by taking the gradient of (\ref{eq:Like_incom1}) $\bigtriangledown_{\theta}$ of $\log ( L({\theta}|\textbf{y}) )$ to find the critical points \cite{Dempster77} 
\begin{equation} 
	\label{eq:Like_incom3}
	\begin{split}
	\bigtriangledown_{\theta} \log  L({\theta}|\textbf{y})  &= \bigtriangledown_{\theta} \log p(\textbf{x} | {\theta}) - \bigtriangledown_{\theta} \log p(\textbf{x} | \textbf{y}, {\theta}).
\end{split}
\end{equation}  
Using the expressions (\ref{eq:expfamily_prop}), (\ref{eq:expfamily_incomp_prop}), $\bigtriangledown_{\theta} \log p(\textbf{x} | {\theta})$ and $\bigtriangledown_{\theta} \log p(\textbf{x} | \textbf{y}, {\theta})$ can be substituted as 
\begin{align*} 
	\bigtriangledown_{\theta} \log  L({\theta}|\textbf{y})  &=  \e{}{S(\textbf{X}) | {\theta}} - \e{}{S(\textbf{X}) | \textbf{y}, {\theta}}. 
\end{align*}
$\e{}{S(\textbf{X})}$ and $\e{}{S(\textbf{X}|\textbf{y})}$ presented in (\ref{eq:expfamily_prop}), (\ref{eq:expfamily_incomp_prop}) involves the natural parameters $\boldsymbol{\eta}$, however, it can be reparameterized to the conventional parameters ${\theta}$, refer Definition \ref{df:def2} $\eta_1 = \log \frac{\alpha}{1-\alpha}, \eta_2 = \log \frac{\beta}{1-\beta}$.  \\ 
Evaluating $\bigtriangledown_{\theta} \log ( L({\theta}|\textbf{y}) ) = 0 $ to find critical points 
\begin{align} 
	\label{eq:Like_incom4}
	\e{}{S(\textbf{X}) | {\theta}} &= \e{}{S(\textbf{X}) | \textbf{y}, {\theta}}.
\end{align}
(\ref{eq:Like_incom4}) gives a relation between the conditional and the unconditional expected value of the sufficient statistic. To solve (\ref{eq:Like_incom4}) for ${\theta}$, the knowledge of $p(\textbf{x} | {\theta})$ on the left hand side and $p(\textbf{x} | \textbf{y}, {\theta})$ on the right hand side is required for calculating the $\mathbb{E}$. This is not available, hence the expression in (\ref{eq:Like_incom4}) doesn't lead to a solution. In order to find a solution, an iterative approach is required. 

%%%%%%%%%%%%%%%%%%%%%%%%%%%%%%%%%%%%%%%%%%%%%%%%%%%%%%%%%%%%%%%%%%%%%%%%%%%%%%%%%%%%%%%%%
\subsection{Expectation and Maximization Algorithm} \label{subsec:EM}
The Expectation and maximization algorithm (E-M algorithm) is an iterative approach $(p = 1,2...,P)$, first proposed by Dempster \textit{et al.} \cite{Dempster77}, that determines the MLE for the incomplete dataset, where the analytic solution given by (\ref{eq:Like_incom4}) doesn't exist. The iteration breaks (\ref{eq:Like_incom4}) into the expectation (E-step) and maximization (M-step) step. Here we apply the E-M algorithm to the Markov process with incomplete dataset.%The iteration is continued until the estimate $\widehat{{\theta}}$ converges. 
\subsubsection{Expectation Step}
The E-step computes the right hand side of (\ref{eq:Like_incom4}) by substituting the $p(\textbf{x} | \textbf{y}, {\theta})$ with unknown ${\theta}$ for $p(\textbf{x} | \textbf{y}, {\theta}^{(p)})$ with known ${\theta}^{(p)}$
\begin{equation}
	\label{eq:EM_expec}
	\begin{bmatrix}
	 	\frac{t_{01}^{(p)}}{{T_0}^{(p)}} & \frac{t_{10}^{(p)}}{{T_1}^{(p)}}.
	\end{bmatrix} = \e{}{\textbf{S}(\textbf{X})|\textbf{y}, \theta^{(p)}} 
\end{equation}
In the $(p)^{th}$ iteration, (\ref{eq:EM_expec}) evaluates the conditional expectation $\e{}{\textbf{S}(\textbf{X})|\textbf{y}, \theta^{(p)}}$ to obtain the statistics $t_{01}^{(p)}, t_{10}^{(p)}, {T_0}^{(p)}, {T_1}^{(p)}$ with known ${\theta}^{(p)}$ and given \textbf{y}, where ${\theta}^{(p)}$ is the value of the parameter determined from the ${(p-1)}^{th}$ step. Following the mathematical intuition, E-step completes the missing information to obtain $t_{01}^{(p)}, t_{10}^{(p)}, {T_0}^{(p)}, {T_1}^{(p)}$ required for estimation.

\subsubsection{Maximization Step} 
The M-step performs the maximization by computing left hand side of (\ref{eq:Like_incom4})
\begin{equation}
	\begin{split}
	\label{eq:EM_maximize1}	
		\e{}{\textbf{S}(\textbf{X})| \theta^{(p+1)}} &=
	\begin{bmatrix}
	 	\frac{t_{01}^{(p)}}{{T_0}^{(p)}} & \frac{t_{10}^{(p)}}{{T_1}^{(p)}}
	\end{bmatrix}, 
	\end{split}
\end{equation}
i.e., the statistics $t_{01}^{(p)}, t_{10}^{(p)}, {T_0}^{(p)}, {T_1}^{(p)}$ evaluated in the E-step are used to determine the estimates $\widehat{\theta}^{(p + 1)}$ for the ${(p+1)}^{th}$  E-step
\begin{equation}
	\begin{split}
	\label{eq:EM_maximize2}	
	\begin{bmatrix}
	 		 \widehat{\alpha}^{(p + 1)} & \widehat{\beta}^{(p + 1)}
	\end{bmatrix}  &=
	\begin{bmatrix}
	 	\frac{t_{01}^{(p)}}{{T_0}^{(p)}} & \frac{t_{10}^{(p)}}{{T_1}^{(p)}}
	\end{bmatrix}.
	\end{split}
\end{equation}    
The convergence is obtained at a critical point where $({\widehat{\alpha}}^{(p)}, {\widehat{\beta}}^{(p)})$ satisfies (\ref{eq:Like_incom4}).   \\
For $p=1$, the E-M algorithm requires to initialize the parameters to a starting value $(\widehat{\alpha}^{(0)}, \widehat{\beta}^{(0)})$. Following the definition of the incomplete dataset from (\ref{eq:Incompl_Representation}), the likelihood function has multiple modes. Therefore, it is important to start with different choices of starting values. Upon convergence of estimates to their local maximum, estimate with the least squared error is selected as the global maximum.\\ %Each iteration of an E-M algorithm requires to compute  $p(\textbf{x} | \textbf{y}, {\theta}^{(p)})$, followed by $\mathbb{E}$.
To find the global maximum, the squared error
\begin{equation}
	\text{SE} = 10 \cdot \log \left( |p(\textbf{y}|(\widehat{\alpha}^{(p)}, \widehat{\beta}^{(p)})) - p(\textbf{y}|(\alpha, \beta))|^2 \right)
\end{equation}
is computed at local maxima for each starting choice. $p(\textbf{y}|(\alpha, \beta))$ described in (\ref{eq:Like_incom_probrel_ex}) is evaluated after observing $\textbf{y}$, hence known at SU. $p(\textbf{y}|(\widehat{\alpha}^{(p)}, \widehat{\beta}^{(p)}))$ is determined at local maxima $(\widehat{\alpha}^{(p)}, \widehat{\beta}^{(p)})$.
\begin{figure*}[!t]
\subfloat{\includegraphics[trim=0.0cm 0.0cm 0.0cm 0.0cm,clip=true,scale = 1]{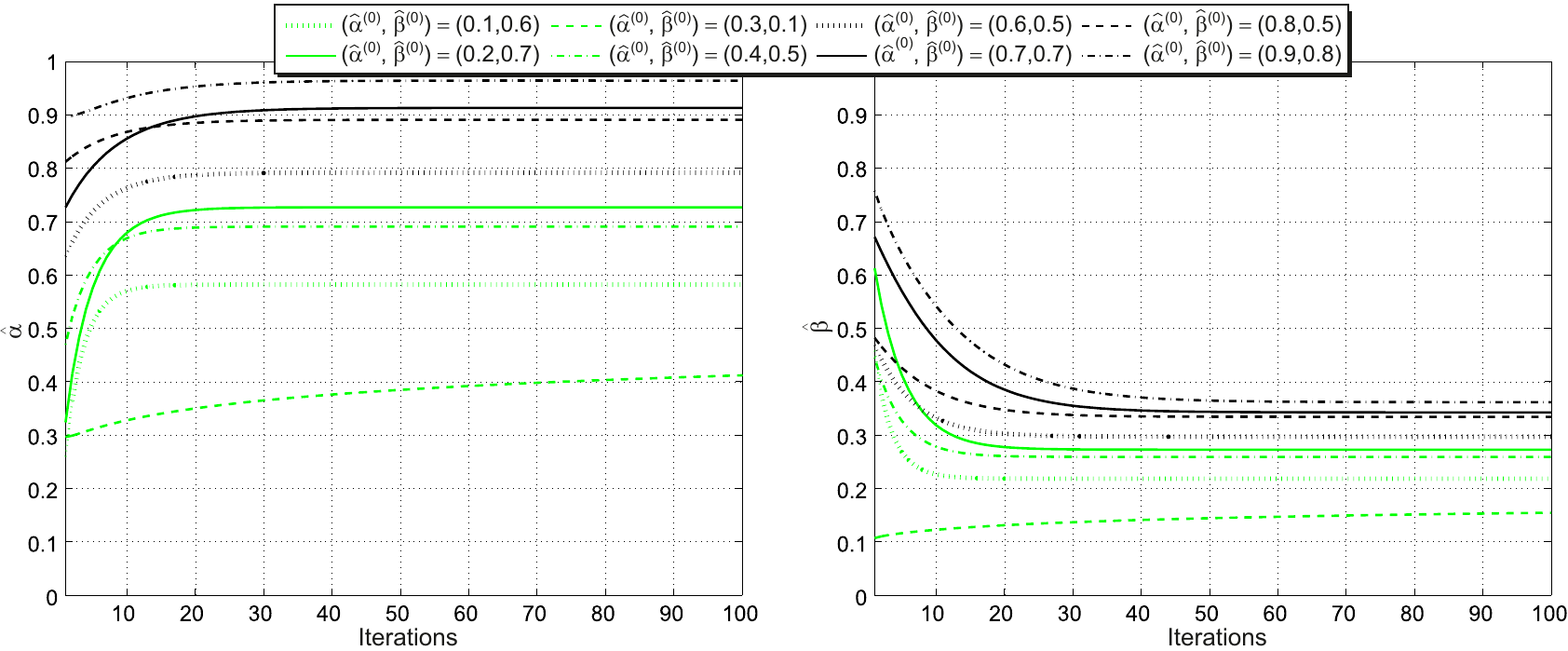}}
\caption{The convergence of the estimates $(\widehat{\alpha}^{(p)}, \widehat{\beta}^{(p)})$ against the iterations following the E-M algorithm where different choices of starting values $(\alpha^{(0)}, \beta^{(0)})$ are considered. }
\label{fig:TransVsIt}
\end{figure*}
%%%%%%%%%%%%%%%%%%%%%%%%%%%%%%%%%%%%%%%%%%%%%%%%%%%%%%%%%%%%%%%%%%%%%%%%%%%%%%%%%%%%%%%%%
\section{Simulation Results} \label{sec:sim res}
%%%%%%%%%%%%%%%%%%%%%%%%%%%%%%%%%%%%%%%%%%%%%%%%%%%%%%%%%%%%%%%%%%%%%%%%%%%%%%%%%%%%%%%%%
To check the validity of the E-M algorithm, Monte Carlo simulations are performed. The slot sequence for a single channel is generated using Markov process with known $(\alpha, \beta)$. To realize incomplete dataset, unobserved slots according to $L$ are declared undefined.  \\
The SU implements E-M algorithm over the observed slots to determine the estimates $(\widehat{\alpha}, \widehat{\beta})$. The iterations for the E-M algorithm are carried out over the software. For analysis, the intermediate values $\widehat{\theta}^{(p)}$ for $p^{th}$ iteration are stored. To obtain a considerable amount of estimation accuracy for the parameters, the number of slots observed by SU is fixed to $K = 10^6$.\\ 
Consider a single realization of a channel with $L=4$, i.e., SU observes every $5^{th}$ slot of the channel. $(\alpha, \beta) = (0.8, 0.3)$ are chosen to be the true value of the parameters. \figurename \ref{fig:TransVsIt} illustrates the convergence of the estimates $(\widehat{\alpha}^{(p)}, \widehat{\beta}^{(p)})$ with the number of iterations of the E-M algorithm. More than 80 $\%$ of the starting values attain their local maximum within 50 iterations. It is seen that starting value $(\widehat{\alpha}^{(0)}, \widehat{\beta}^{(0)}) =(0.6,0.5)$ converges to its local maximum $(0.791,0.297)$ at $p = 30$, which is closest to the true value (0.8, 0.3). \\ 
\begin{table}[!t]
\renewcommand{\arraystretch}{1.3}
\caption{Convergence and Squared Error at p = 100 for different choices of starting values}
\label{tb:SE}
\centering
\begin{tabular}{c||c|c}
\hline
\bfseries $(\widehat{\alpha}^{(0)}, \widehat{\beta}^{(0)})$  & $(\widehat{\alpha}^{(100)}, \widehat{\beta}^{(100)})$  & SE (dB)\\
\hline\hline
(0.1,0.6) & (0.581, 0.218) & -66.6 \\ \hline
(0.2,0.7) & (0.726, 0.272) & -96.9 \\ \hline
(0.3,0.1) & (0.411, 0.154) & -45.9 \\ \hline
(0.4,0.5) & (0.690, 0.259) & -96.7 \\ \hline
(0.6,0.5) & (0.791, 0.297) & -103.7 \\ \hline 
(0.7,0.7) & (0.913, 0.343) & -56.7 \\ \hline
(0.8,0.5) & (0.890, 0.334) & -62.1 \\ \hline
(0.9,0.8) & (0.964, 0.361) & -46.8 \\ \hline
\end{tabular}
\end{table}
\begin{figure}[!t]
	\centering
 	\includegraphics[trim=0.0cm 0.0cm 0.0cm 0.0cm,clip=true,width=\columnwidth]{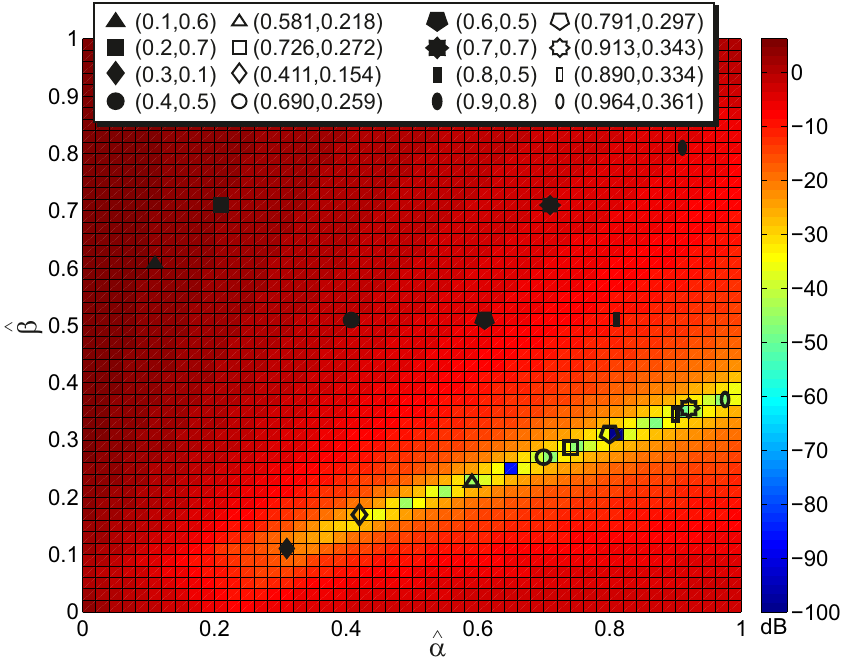}
	\caption{Squared error (dB) along the parameter scale $(\widehat{\alpha}, \widehat{\beta}) \in [0,1]$. The different starting values $(\widehat{\alpha}^{(0)}, \widehat{\beta}^{(0)})$ and their values $(\widehat{\alpha}^{(100)}, \widehat{\beta}^{(100)})$ after $100^{th}$ iteration of the E-M algorithm. $(\alpha, \beta) = (0.8, 0.3)$ is the true value of the parameters.}
	\label{fig:SquaredError}
\end{figure}
To determine the global maximum, SE for the convergence points are computed. Table \ref{tb:SE} provides the SE for the local maxima at $(\widehat{\alpha}^{(100)}, \widehat{\beta}^{(100)})$ for the different starting choices. To illustrate the correlation between SE and modes for the function $p(\textbf{y}|(\alpha, \beta))$, SE is computed for $(\widehat{\alpha}, \widehat{\beta}) \in [0,1]$ at steps of $0.02$ and plotted as a color map in \figurename \ref{fig:SquaredError}. Please notice, \figurename \ref{fig:SquaredError} intends to demonstrate the convergence of the E-M algorithm at local maxima. However limited by its resolution, it doesn't represent all the local maximum. 
To follow the convergence path of the estimates obtained from E-M, the starting choices $(\widehat{\alpha}^{(0)}, \widehat{\beta}^{(0)})$ and their values $(\widehat{\alpha}^{(100)}, \widehat{\beta}^{(100)})$ after $100^{th}$ iteration are mapped into the color map. \\
\figurename \ref{fig:SquaredError} shows that for all starting values $(\widehat{\alpha}^{(0)}, \widehat{\beta}^{(0)})$, the E-M algorithm converges to the local maximum that also has the lowest SE in the region. Table \ref{tb:SE} and \figurename \ref{fig:SquaredError} validate that the value $(\widehat{\alpha}^{(0)}, \widehat{\beta}^{(0)}) = (0.6,0.5)$ attains its convergence at $(\widehat{\alpha}^{(100)}, \widehat{\beta}^{(100)}) = (0.791,0.297)$, that has the least SE among the starting choices, is the global maximum. Hence, is the estimate of the parameters. The number of modes in the likelihood function is proportional to $L$. Large $L$, i.e., $L > 10$ corresponds to a considerable increase in the number of starting choices to determine the global maximum. \\
\begin{figure}[!t]
	\centering
\includegraphics[trim=0.0cm 0.0cm 0.0cm 0.0cm,clip=true,width=\columnwidth]{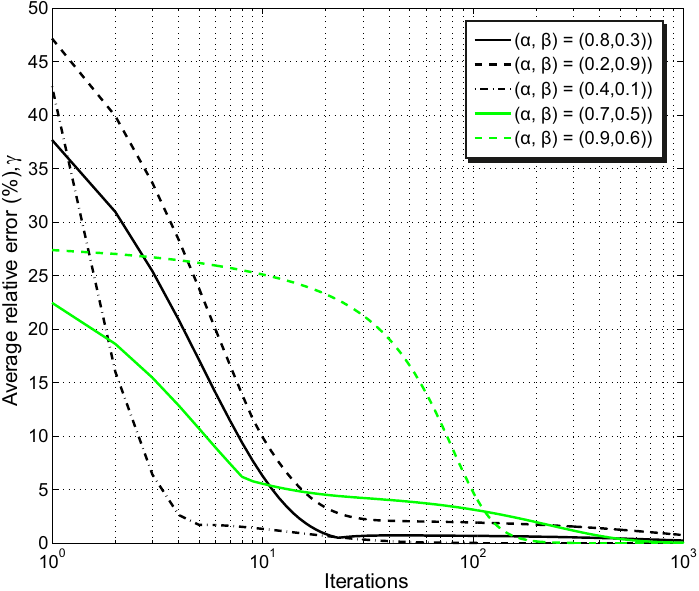}
	\caption{Average relative error against the number of iterations for the E-M algorithm for 5 PU channels with random skipped slot length $L \in \{1,2,3,4,5,6\}$. The transitional probabilities for the channels $(\alpha, \beta) = [(0.8,0.3),(0.2,0.9),(0.4,0.1),(0.7,0.5),(0.9,0.6) ]$ with starting choice $(\widehat{\alpha}^{(0)}, \widehat{\beta}^{(0)}) = [(0.6,0.5),(0.4,0.8),(0.1,0.3),(0.8,0.3),(0.7,0.4) ]$ and utilization probabilities $\textbf{u} = [0.27, 0.81, 0.2, 0.41, 0.4]$ evaluated using (\ref{eq:chauti2}).}
	\label{fig:RelaterrVsIt}
\end{figure}
The convergence accuracy for the E-M algorithm is examined using average relative error $\gamma$, defined as
\begin{equation}
\gamma = \frac{1}{2} \left( \frac{| \alpha - \widehat{\alpha}^{(p)} |}{\alpha} + \frac{|\beta - \widehat{\beta}^{(p)}|}{\beta} \right) \cdot 100.
\end{equation}
\figurename \ref{fig:RelaterrVsIt} considers the validation of E-M algorithm to a real scenario, where SU observes $N = 5$ PU channels, each having a different $(\alpha, \beta)$ and $L \in \{1,2,3,4,5,6\}$, i.e., each channel is observed by the SU after a skipping $L$ slots, where $L$ is uniformly distributed over $\{1,2,3,4,5,6\}$. For the starting choice of $(\widehat{\alpha}^{(0)}, \widehat{\beta}^{(0)})$ for each channel, only the value with least SE is examined. \\ 
For an appropriate of $(\widehat{\alpha}^{(0)}, \widehat{\beta}^{(0)})$, E-M algorithm estimates the parameters with $\gamma < 5\%$, for all the channels after $p = 20$ iterations and $\gamma < 1\%$ for $80 \%$, of the channels after $p = 1000$.  %Increasing $L$ has an influence on the number of modes for the likelihood function, therefore for optimum convergence a large number of starting values are required. 
$(\alpha, \beta)$ determine the 2-D slope of the likelihood function. For its fixed value of $(\alpha, \beta)$, the rate of convergence depends on $L$ and the starting choice $(\widehat{\alpha}^{(0)}, \widehat{\beta}^{(0)})$.  \\ %Finally, applying E-M algorithm to a general case $L = \{3,5,7\}$, i.e., the SU observes the channel by selecting the $L$ randomly (uniform distribution) between $3,5,7$. For the case $(0.8,0.3)$ the convergence rate for $L = \{3,5,7\}$ lies between $L = 4$ and $L = 6$ and for $(0.2,0.9)$ convergence rate for $L = \{3,5,7\}$ is similar to $L = 4$.   
Not considered in the paper, however several heuristic approaches can be applied for choosing $(\widehat{\alpha}^{(0)}, \widehat{\beta}^{(0)})$, for example, $T_0$ and $T_1$ are evaluated from (\ref{eq:Suff1}) after observing $\textbf{y}$, thereby $u$ is determined using $\frac{1-u}{u} = \frac{T_1}{T_0}$ from (\ref{eq:chauti1}). Inserting the value $u$ in (\ref{eq:chauti2}), the slope $m = \frac{\beta}{\alpha} = \frac{1-u}{u}$ can be obtained. This states that local maxima exist across the points $(\eta, m \cdot \eta)$ where $\eta \in [0,1]$, see \figurename \ref{fig:SquaredError} where $m = \frac{0.3}{0.8}$. Choosing a value close to $(\eta,m \cdot \eta)$ increases the convergence speed of the E-M algorithm. \\
The E-M algorithm is an iterative approach based on MLE, that determines the critical points for the likelihood function. The MLE properties, i.e., unbiased and minimum variance are however retained. 
   
%%%%%%%%%%%%%%%%%%%%%%%%%%%%%%%%%%%%%%%%%%%%%%%%%%%%%%%%%%%%%%%%%%%%%%%%%%%%%%%%%%%%%%%%%
\section{Conclusion} \label{sec:conc}
%%%%%%%%%%%%%%%%%%%%%%%%%%%%%%%%%%%%%%%%%%%%%%%%%%%%%%%%%%%%%%%%%%%%%%%%%%%%%%%%%%%%%%%%%
%The paper describes SU as a cognitive radio entity that finds spectrum holes inside the PU spectrum. 
The model parameters, i.e, state transition probabilities enable the SU to implement reinforcement learning and utilize spectrum holes efficiently. The model parameters are not known at the SU and need to be estimated. However, sensing all PU channels in every slot is not possible making the estimation of the parameters difficult.
%The E-M algorithm provides an optimum solution by determining the Maximum Likelihood estimates when the slots are sensed non-consecutively. 
In this paper, we derived an analytical expression of the likelihood function for the Markov process for the complete and incomplete dataset. The paper also proposes the E-M algorithm to estimate the parameters when the slots are sensed non-consecutively. The simulations show the validity of the estimates obtained using E-M algorithm. To find the global maximum, the least squared error approach is applied. The paper considers perfect sensing, the effect on estimation due to presence of noise and channel will be considered in the future work. 

%%%%%%%%%%%%%%%%%%%%%%%%%%%%%%%%%%%%%%%%%%%%%%%%%%%%%%%%%%%%%%%%%%%%%%%%%%%%%%%%%%%%%%%%%
\section*{Appendix} \label{sec:appen}
%%%%%%%%%%%%%%%%%%%%%%%%%%%%%%%%%%%%%%%%%%%%%%%%%%%%%%%%%%%%%%%%%%%%%%%%%%%%%%%%%%%%%%%
The E-M algorithm is a powerful tool to determine MLE for an incomplete dataset. However, it is applicable only where the underlying distribution belongs to a class of exponential families. To confirm its application over the Markov process, it is important to describe Markov process as an exponential family. 

% \begin{figure*}[!t]	
% ensure that we have normalsize text
% \normalsize
% Store the current equation number.
%  \setcounter{MYtempeqncnt}{\value{equation}}
% Set the equation number to one less than the one
% desired for the fi	rst equation here.
% The value here will have to changed if equations
% are added or removed prior to the place these
% equations are referenced in the main text.
%\setcounter{equation}{5}
\begin{defi}[Exponential family]
\label{df:def1}
The probability mass function of a random vector $\textbf{X} = [X_1,X_2,...,X_t]$ with dimension $t$ belongs to an exponential family if it can be represented in the form \cite{kobayashi}	
	\begin{equation}
		\label{eq:expfamily}
		p(\textbf{x}|\boldsymbol{\eta}) = h(\textbf{x}) \cdot e^{{\boldsymbol{\eta}}' \cdot \textbf{S}(\textbf{x}) - A(\boldsymbol{\eta})},
		\end{equation}
where $\boldsymbol{\theta}) = [\theta_1, \theta_2,....,\theta_d] \in \Theta \subset {\mathbb{R}}^d$ are the convectional parameters. $\textbf{x} \in \mathcal{X}^t \subset \mathbb{R}^t$. $\boldsymbol{\eta} = [\eta_1,\eta_2,..., \eta_m]$ are the natural or the canonical parameters. $\textbf{S} = [S_1,S_2,..., S_m]$ are the sufficient statistics. Each $\eta_i$ corresponds to a mapping from the conventional parameter space to the canonical parameter space, i.e., $\eta_i: \Theta \rightarrow \mathcal{N} \subset \mathbb{R}$. The $i^{th}$ sufficient statistic $S_i: \mathcal{X}^t \rightarrow \mathbb{R}$ for $i = 1,2,..m$.  
$h:\mathcal{X} \rightarrow {\mathbb{R}}_{>0}$ is the support function. $A(\boldsymbol{\eta}): \mathcal{N}^m \rightarrow {\mathbb{R}}_{>0}$ is the normalizing function or the cumulant moment generating function for $\textbf{S}(\textbf{x})$. Finally, $m,t,d \in \mathbb{N}$. $(\cdot)'$ represents the matrix transpose. \\ 
Reformulating (\ref{eq:expfamily}) to represent $A(\boldsymbol{\eta})$ in terms of $\textbf{S}(\textbf{x})$ and $h(\textbf{x})$ and inserting $t = T$ gives
	\begin{equation}
		\label{eq:expfamily_mgf}
		A(\boldsymbol{\eta}) = \log \int\limits_{\textbf{x} \in \mathcal{X}^T} h(\textbf{x}) \cdot e^{{\boldsymbol{\eta}}' \cdot \textbf{S}(\textbf{x}) } d\textbf{x}. 
	\end{equation}
Clearly, $A(\boldsymbol{\eta})$ in (\ref{eq:expfamily_mgf}) represents the cumulant moment generating function for $\textbf{S}(\textbf{x})$, solving
\begin{align}
	\label{eq:expfamily_prop}	
	\bigtriangledown_{\boldsymbol{\eta}} A(\boldsymbol{\eta}) &= \e{}{\textbf{S}(\textbf{X})},
\end{align}
determines the critical points of the maximum likelihood estimates.\\
Assuming ergodicity, the $\mathbb{E}$ is replaced by sample sum $\sum$, for i.i.d. sequence $\textbf{x} = [x_1, x_2,...x_T] $, it follows that
\begin{equation}
	\label{eq:expfamily_prop_approx}
		\bigtriangledown_{\boldsymbol{\eta}} A(\widehat{\boldsymbol{\eta}}) = \frac{1}{T} \sum_{t}^{T}{\textbf{S}(x_t)}.			
\end{equation}	
Following (\ref{eq:expfamily_mgf}), the conditional probability measure $p(\textbf{x} | \textbf{y}, \boldsymbol{\eta})$ and $A(\boldsymbol{\eta} | \textbf{y})$ are determined as
\begin{align}
	\label{eq:expfamily_incomp}
	p(\textbf{x}|\textbf{y},\boldsymbol{\eta}) &= h(\textbf{x}) \cdot e^{{\boldsymbol{\eta}}' \cdot \textbf{S}(\textbf{x}) - A(\boldsymbol{\eta} | \textbf{y})}, \\
	A(\boldsymbol{\eta}|\textbf{y}) &= \log \int\limits_{\textbf{x} \in \mathcal{X}^T(\textbf{y})} h(\textbf{x}) \cdot e^{{\boldsymbol{\eta}}' \cdot \textbf{S}(\textbf{x}) } d\textbf{x}, \text{  gives} 
\end{align}
\begin{align}
	\label{eq:expfamily_incomp_prop}
	\bigtriangledown_{\boldsymbol{\eta}} A(\boldsymbol{\eta}|\textbf{y}) &= \e{}{\textbf{S}(\textbf{X})|\textbf{y}}.
\end{align}
\end{defi}
\begin{defi}[Maximum likelihood estimation for the Markov process]
\label{df:def2}
For a sequence, $\textbf{X} = [X_1,X_2,...,X_t,..,X_T]$ are related through the Markov property. The pairs $(X_t ,X_{t-1})$ and $(X_{t + 1} ,X_{t})$ are conditionally independent, hence i.i.d. Using this property, the Markov process can be represented as mixture of two exponential families, where $d = 1, m = 1, t = T$.
\begin{align}
	\label{eq:expfamily_Markov}
			p( \textbf{x} | {\theta} ) &=  p(x_{1}|{\theta}) \cdot \prod_{t=2}^{T} p(x_{t + 1}|x_{t}, {\theta}) \nonumber \\ 
			\quad &=  
			 (1 - \alpha)^{\sum_{t=2}^{T} (1 - \mathbbm{1}_{ \{x_t = 1 | x_{t-1} = 0 \} })} \cdot  \alpha^{\sum_{t=2}^{T} \mathbbm{1}_{ \{ x_t = 1 | x_{t-1} = 0 \} }}  \cdot	\nonumber \\
			\quad &  \beta^{\sum_{t=2}^{T} \mathbbm{1}_{ \{ x_t = 0 | x_{t-1} = 1 \} }} \cdot (1 - \beta)^{ \sum_{t=2}^{T} (1 - \mathbbm{1}_{ \{x_t = 0 | x_{t-1} = 1  \} })} \nonumber \\
			\quad &= h_1(\textbf{x}) \cdot e^{ \sum_{t=2}^{T} {\mathbbm{1}}_{ \{ x_t = 1 | x_{t-1} = 0 \} } \cdot \log{ \left( \frac{\alpha}{1-\alpha} \right) } + \log (1-\alpha) } \cdot \nonumber \\
			\quad &
			h_2(\textbf{x}) \cdot e^{ \sum_{t=2}^{T} \mathbbm{1}_{ \{ x_t = 0 | x_{t-1} = 1 \} } \cdot \log{ \left( \frac{\beta}{1-\beta} \right) } +  \log (1-\beta)} 
\end{align}
$(\alpha, \beta) \in [0,1], (x_t, x_{t-1}) \in \{0,1\}^2$, $ \eta_1(\alpha) = \log{ \left( \frac{\alpha}{1-\alpha} \right) }$,  
$ \eta_2(\beta) = \log{ \left( \frac{\beta}{1-\beta} \right) }$, $S_1(x_t,x_{t-1}) = {\mathbbm{1}}_{ \{ x_t = 1 | x_{t-1} = 0 \} }$, $S_2(x_t,x_{t-1}) = {\mathbbm{1}}_{ \{ x_t = 0 | x_{t-1} = 1 \}}$, $A_1 = \log (1+e^{\eta_1}),$ $A_2 =  \log (1+e^{\eta_2})$, \\ \newpage $h_1(\textbf{x}) = \prod_{t=2}^{T} {\mathbbm{1}}_{ \{ x_t \in \{ 0,1 \} \cup x_{t-1}  = 0 \} }$, $h_2(\textbf{x}) = \prod_{t=2}^{T} {\mathbbm{1}}_{ \{ x_t \in \{ 0,1 \} \cup x_{t-1}  = 1 \} }$. \\ 
Following the result in (\ref{eq:expfamily_prop}), the underlying Markov process defined in (\ref{eq:expfamily_Markov}) is represented as
\begin{align}
	\label{eq:expfamily_prop_Markov}	
	\begin{bmatrix} 
			\frac{d}{d{\eta}_1} A_1({\eta}_1) & \frac{d}{d{\eta}_2} A_2({\eta}_2)	
		\end{bmatrix}	
		 &= 
		 \begin{bmatrix}
		 \e{}{S_1(\textbf{X})} & \e{}{S_2(\textbf{X})}
  	 \end{bmatrix}.
\end{align}
Using the conditional independence between $(X_t ,X_{t-1})$ and $(X_{t + 1} ,X_{t})$, and applying ergodicity gives 
	\begin{align}
	\label{eq:expfamily_Markov_ergodic}
		\begin{bmatrix} 
			 \frac{d}{d{\eta}_1} A_1(\widehat{{\eta}}_1) & \frac{d}{d{\eta}_2} A_2(\widehat{{\eta}}_2)	
		\end{bmatrix} = \hspace*{5cm} \nonumber \\	  
		\begin{bmatrix} \frac{1}{T_0} \sum\limits_{t=2}^{T} \mathbbm{1}_{ \{ x_t = 1 | x_{t-1} = 0  \}}  & \frac{1}{T_1} \sum\limits_{t=2}^{T} \mathbbm{1}_{ \{ x_t = 0 | x_{t-1} = 1  \}} 
		\end{bmatrix}, 		
	\end{align}			  
where $T_{0} \eqdef \sum_{t}^{T} \mathbbm{1}_{ \{ x_{t} | x_{t - 1} = 0 \} }$ and $T_{1} \eqdef \sum_{t}^{T} \mathbbm{1}_{ \{ x_{t} | x_{t - 1} = 1 \} }$.		
\end{defi}
% Restore the current equation number.
% IEEE uses as a separator
%\hrulefill
% The spacer can be tweaked to stop underfull vboxes.
%\vspace*{0.1pt}
%\end{figure*}
%%%%%%%%%%%%%%%%%%%%%%%%%%%%%%%%%%%%%%%%%%%%%%%%%%%%%%%%%%%%%%%%%%%%%%%%%%%%%%%%%%%%%%%%%
% References
%%%%%%%%%%%%%%%%%%%%%%%%%%%%%%%%%%%%%%%%%%%%%%%%%%%%%%%%%%%%%%%%%%%%%%%%%%%%%%%%%%%%%%%%%
%\newpage
\bibliographystyle{IEEEtran}
\bibliography{IEEEabrv,refs}

\end{document}